\documentclass[12pt]{article}
\usepackage{amsmath}
\usepackage{amssymb}
\usepackage{graphicx}
\usepackage{cite}

\makeatletter
\usepackage{indentfirst}
\usepackage{bm}
\usepackage{epsfig}

\makeatother

\begin{document}

%-------------------  First Head  -----------------------------------------
% \thispagestyle{empty} \vspace*{0.8cm}
% \hbox to\textwidth{\vbox{\hfill{}\textsf{\huge{}Commun. Theor.
% Phys.\hfill{}}}}{\huge\par}

% \noindent \rule[3mm]{1\textwidth}{0.2pt}\hspace*{-1\textwidth}\rule[2.5mm]{1\textwidth}{0.2pt}

% \noindent %=================== Text begin here =============================================
\begin{center}
\textbf{\LARGE{}Determining the molecular Huang-Rhys factor via STM
induced luminescence}{\LARGE\par}
\par\end{center}

\textbf{\LARGE{}\footnotetext{\hspace*{-0.45cm}}\textbf{\footnotesize{}$
% ^{*}...........
$}
\footnotetext{\hspace*{-0.45cm}$^{\dag}$Corresponding author, E-mail:
20210076@sicnu.edu.cn }}{\footnotesize\par}
\begin{center}
{\footnotesize{}Fei Wen$^{{\rm a)}}$ and  Guohui Dong$^{{\rm b)\dagger}}$}{\footnotesize\par}
\par\end{center}

\begin{center}
{\footnotesize{} \sl $^{{\rm a)}}$Graduate
School of China Academy of Engineering Physics, Beijing 100084, China
}\\
{\footnotesize{} $^{{\rm b)}}$ College of Physics and Electronic Engineering,
Sichuan Normal University, Chengdu 610068, China}\\
{\footnotesize{} }{\footnotesize\par}
\par\end{center}

% \begin{center}
% {\footnotesize{}(Received XXXX; revised manuscript received XXXX)}{\footnotesize\par}
% \par\end{center}

{\footnotesize{}\vspace*{2mm}
}{\footnotesize\par}
\begin{center}
{\footnotesize{}}%
\begin{minipage}[c]{15.5cm}%
\indent {\footnotesize{}The scanning tunneling microscopy induced
luminescence (STML) can be used to probe the optical and electronic
properties of molecules. Concerning the vibronic coupling, we model the molecule as a two-level system with the vibrational degrees of freedom.
Based on the Bardeen's theory, we express the inelastic tunneling current in terms of Huang-Rhys factor within the inelastic electron
scattering (IES) mechanism. We find that the differential conductance, varying
with the bias voltage, exhibits distinct step structure with various
vibronic coupling strength. The second derivative of the inelastic tunneling
current with respect to the bias voltage shows the characteristics
of vibrational-level structure with Franck-Condon factor. Consequently,
we propose a method to determine the Huang-Rhys factor of molecules,
holding promising potential within the realm of solid-state physics.}%
\end{minipage}{\footnotesize\par}
\par\end{center}

\begin{center}
{\footnotesize{}}%
\begin{minipage}[c]{15.5cm}%
\begin{minipage}[t]{2.3cm}%
\textbf{Keywords:}%
\end{minipage}%
\begin{minipage}[t]{13.1cm}%
scanning tunneling microscopy induced luminescences (STML), Franck-Condon
factor, Huang-Rhys factor, inelastic electron scattering, differential
conductance%
\end{minipage}

\vglue8pt%
\end{minipage}{\footnotesize\par}
\par\end{center}

\section{Introduction}

Scanning tunneling microscopy induced luminescence (STML) serves as
a powerful tool for detecting various molecular properties, including
molecular conformation\cite{Qiu2003,Schwarz_2015} and spectral characteristics
\cite{Qiu2003,Schwarz_2015,Qiu_2004,Wu_2008,Chen_2010,Zhang_2016}.
Researchers have made substantial progresses in understanding the
fine structures of molecular energy by combining the current-voltage
characteristics of molecular junctions with optical spectroscopy \cite{Komeda_2005,Katano_2010,Jiang_2012,Zhang_2017,Miwa_2019,Kong_2021,Pe_a_Rom_n_2022,Wen_2022}. Intramolecular transitions with vibronic features have also been successfully detected in the differential conductance spectra at the threshold
of a vibrational mode energy \cite{Schwarz_2015,Stipe1998,Stipe_1998,Lauhon_1999,Lorente_2000,Pradhan_2005,Nazin_2005}.

In 1950, Huang and co-workers derived the expression for the Huang-Rhys
factor (denoted as $S$ ) which characterizes the strength of the
vibronic coupling\cite{Huang1950}. The critical importance of the
Huang-Rhys factor was confirmed in both theory and experiments\cite{Lemos_1965,MULAZZI_1967,Moreno1992,Schenk_1992}.
Although a simple model which assumes that  all vibration modes
have the same frequency under the Born-Oppenheimer approximation is used.  In many practical applications, the
results derived from this simplest model have been proven to be the most useful. There are several ways to determine the Huang-Rhys
factor in experiments, such as fitting the spectrum and extracting
$S$ from the Stokes shift. However, these methods often require accurate
spectral shapes and higher resolution, which can be challenging to
achieve in practice\cite{Jong2015}. Taking advantage of the low temperature
and excellent accuracy of STML, measuring the Huang-Rhys factor becomes
a more accessible task.

In general, the luminescence is induced by STM with three mechanisms:
the inelastic electron scattering (IES) mechanism \cite{Komeda_2005,Mingo_2000,Tikhodeev_2001,Mii_2002,Hurley_2011,Eickhoff_2020,Dong_2020,Dong_2021}, charge injection (CI) mechanism \cite{Schwarz_2015,Miwa_2019,Drakova_1997,Galperin_2005,Harbola_2006,Jiang_2023}
and the plasmon mechanism \cite{Kong_2021,Doppagne_2017,Mart_n_Jim_nez_2020,Zhu_2021,Miwa_2023}. Here, we specifically focus on the IES mechanism, where electrons
tunnel inelastically from one electrode to another, exciting the molecule
in the gap of the tip and the substrate.  Taking the vibronic coupling into account, we calculate the
inelastic tunneling current based on the perturbation theory \cite{Bardeen_1961}.
The inelastic tunneling current is expressed in terms of the Franck-Condon
factor which describes the square of the overlap integral between
the vibrational wave functions of the two states that are involved
in the transition. The Franck-Condon factor is written in the form
of the Huang-Rhys factor \cite{Franck_1926,Condon_1926}, which is
determined through the second derivative of the inelastic tunneling
current with respect to the bias voltage.

The rest of the paper is organized as follows. In section \ref{sec:Model-and-method}, we describe the Hamiltonian of our model in which the molecule is
treated as a two-level system with the vibration degree of the freedom.
Then, we derive the inelastic tunneling current in terms of Huang-Rhys
factor. Section \ref{sec:Results} shows the inelastic tunneling current,
the differential conductance and the second derivative of the inelastic
tunneling current with respect to the bias voltage as a function of
the bias voltage. Finally, we summarize the main contributions in
section \ref{sec:Conclusion}.

\section{Model and method\label{sec:Model-and-method}}

\subsection{Hamiltonian}

The STML system comprises a metallic tip, a molecular sample, a decoupling
layer and a metallic substrate. The decoupling layer serves to separate
the molecule from the substrate, blocking the quenching process and
enabling the molecular luminescence. In the IES mechanism, when a
bias voltage is applied between the tip and substrate, electrons tunnel
from one electrode to another, thereby exciting the molecule.

The total Hamiltonian is divided into three components: the electronic
Hamiltonian, the molecular Hamiltonian and the interaction Hamiltonian.The
electronic component is solved by Bardeen's theory. The molecular
aspect is simplified as a two-level system with vibrational degrees
of freedom. The interaction between the electron and the molecule
is Coulomb interaction.

Based on Bardeen's theory, it is assumed that the tip and the substrate
are separated from each other. The stationary Schr{\"o}dinger equations
are
\begin{equation}
\hat{H}_{t}\left|\phi_{k}\right\rangle =\tilde{\xi}_{k}\left|\phi_{k}\right\rangle ,\label{eq:Ht}
\end{equation}
\begin{equation}
\hat{H}_{s}\left|\varphi_{n}\right\rangle =\tilde{E}_{n}\left|\varphi_{n}\right\rangle ,\label{eq:Hs}
\end{equation}
where $\hat{H}_{t}$ and $\hat{H}_{s}$ represent the Hamiltonians
of electrons in the tip and the substrate respectively. $\phi_{k}$
and $\varphi_{n}$ are the electronic wave function of the corresponding
region. $\tilde{\xi}_{k}$ and $\tilde{E}_{n}$ are the energies of
the tunneling electrons located at the tip and the substrate region
when the bias voltage $V_{b}$ is applied between the electrodes.
$\tilde{\xi}_{k}$ and $\tilde{E}_{n}$ are expressed in terms of
the electronic energy in the absence of the bias voltage,
\begin{equation}
\tilde{\xi}_{k}=\xi_{k}+eV_{b},
\end{equation}
\begin{equation}
\tilde{E}_{n}=E_{n},
\end{equation}
where $\xi_{k}$ $\left(E_{n}\right)$ is the energy of the tunneling
electron at the tip (substrate) when the bias is zero. $e$ represents
the elementary charge of an electron. Selecting the center of the
molecule as the origin of the coordinates, the wave functions in Eq.
\ref{eq:Ht} and Eq. \ref{eq:Hs} are explicitly written as
\begin{equation}
\phi_{k}\left(\vec{r}\right)=A_{k}\frac{e^{-\kappa_{k}\left(\left|\vec{r}-\vec{a}\right|-R\right)}}{\kappa_{k}\left|\vec{r}-\vec{a}\right|},
\end{equation}
\begin{equation}
\varphi_{n}\left(\vec{r}\right)=B_{n}e^{-\kappa_{n}\left|z\right|},
\end{equation}
in which $A_{k}$ and $B_{n}$ are the normalized coefficients. $\vec{r}$
represents the position of the tunneling electron. $\left|z\right|$
denotes the distance between the tunneling electron and the substate.
The apex of the tip is assumed to be sphere.The coordinates of the
center and the radius of curvature are denote as $\vec{a}$ and $R$.
$\kappa_{k}$ and $\kappa_{n}$ are the decay constants and can be
expressed as 
\begin{equation}
\kappa_{k}=\frac{\sqrt{-2m_{e}\xi_{k}}}{\hbar},
\end{equation}
\begin{equation}
\kappa_{n}=\frac{\sqrt{-2m_{e}E_{n}}}{\hbar},
\end{equation}
in which $m_{e}$ represents the mass of the electron and $\hbar$
is the reduced Plank constant.

The molecule is modeled as a two-level system with the vibrational
degrees of freedom. The molecular Hamiltonian is 
\begin{equation}
\hat{H}_{m}=\left(E_{g}+\nu_{g}\hbar\omega\right)\left|g,\nu\right\rangle \left\langle g,\nu\right|+\left(E_{e}+\nu^{\prime}\hbar\omega\right)\left|e,\nu\right\rangle \left\langle e,\nu^{\prime}\right|,
\end{equation}
where $\omega$ is the frequency of vibration. $E_{g}$ and $\nu\hbar\omega$
$\left(E_{e}\;\mathrm{and}\;\nu^{\prime}\hbar\omega\right)$ represent
the electronic energy and the vibrational energy associated with the
ground (excited) state of the molecular. $\left|g\right\rangle $
represents the ground state of the molecular electron, while $\left|e\right\rangle $
corresponds to the excited state. $\left|\nu\right\rangle $ ($\left|\nu^{\prime}\right\rangle $)
is the vibrational state with the ground (excited) state of the molecule.

The interaction is approximated as the electron-dipole interaction\cite{Dong_2020},
\begin{equation}
\hat{H}_{el-m}=-e\frac{\vec{r}\cdot\vec{\mu}}{\left|\vec{r}\right|^{3}},
\end{equation}
where $\vec{\mu}$ is the molecular dipole.

\subsection{The Franck-Condon factor}

Within the Born-Oppenheimer approximation, the total wave function
of the molecule is expressed as the product of the electronic wave
function and the vibrational wave function. For the treatment of molecular
vibrations, the the harmonic approximation is often employed. The
Hamiltonian of the nuclear vibration takes the form of a harmonic
oscillator \cite{Huang1950} with the characteristic frequency
$\omega$. The ground states of vibrational states $\left|\nu\right\rangle $
and $\left|\nu^{\prime}\right\rangle $ can be expressed as $\left|0_{g}\right\rangle $
and $\left|0_{e}\right\rangle $, respectively. The subscript $g$
and $e$ indicate the electronic ground state and excited state that
couple with vibrational states. In the molecular excited state, the
harmonic oscillation is viewed as a translation of the harmonic oscillator
when the molecule is in the ground state, thus we establish the following
relationship:
\begin{equation}
\left|0_{e}\right\rangle =\hat{D}\left(\alpha\right)\left|0_{g}\right\rangle ,
\end{equation}
in which $\hat{D}\left(\alpha\right)$ is the translation operator,
capable of translating the harmonic oscillator by a distance of $q$
in real space.
\begin{equation}
q=\alpha\sqrt{\frac{2\hbar}{m_{n}\omega}},
\end{equation}
where $m_{n}$ is the mass of the harmonic oscillation. The Franck-Condon
factor describes the overlap of the vibrational wave function in the
molecular ground state and the molecular excited state. It is expressed
as follows:
\begin{equation}
F_{\nu\nu^{\prime}}=\left|\left\langle \nu\right|\nu^{\prime}\rangle\right|^{2}.\label{eq:FC factor}
\end{equation}
Due to the low temperature of the experiment, the vibrational state
associated with the electronic ground state is assumed to be in its
ground state, i.e., $\left|\nu\right\rangle =\left|0_{g}\right\rangle $.
Then, the Franck-Condon factor is simplified as 
\begin{equation}
\left|\left\langle 0_{g}\right|\nu^{\prime}\rangle\right|^{2}=e^{-S}\frac{S^{\nu^{\prime}}}{\nu^{\prime}!},
\end{equation}
where $S$ is the Huang-Rhys factor and it can be expressed as
\begin{equation}
S=\left|\alpha\right|^{2}=q^{2}\frac{m_{n}\omega}{2\hbar}.\label{eq:S}
\end{equation}

\subsection{The inelastic tunneling current}

At the initial time, we assume that the tunneling electron is located
in the substrate region and the molecule is in the electronic and
vibrational ground state due to the low temperature in experiments.
We express the state of the system at time $t$ as 
\[
\left|\Psi\left(t\right)\right\rangle =e^{-i\left(\tilde{E}_{n}+E_{g}+\nu_{g}\hbar\omega\right)t/\hbar}\left|g,\nu\right\rangle \left|\varphi_{n}\right\rangle +\sum_{k,\nu}c_{g\nu k}\left(t\right)\left|g,\nu\right\rangle \left|\phi_{k}\right\rangle +\sum_{k,\nu^{\prime}}c_{e\nu^{\prime}k}\left(t\right)\left|e,\nu^{\prime}\right\rangle \left|\phi_{k}\right\rangle ,
\]
where $\left|\nu\right\rangle =\left|0_{g}\right\rangle $. $c_{g\nu k}$
represents the elastic tunneling amplitude, while $c_{e\nu^{\prime}k}$
signifies the inelastic tunneling amplitude. Since the molecular vibration
does not contribute to the elastic current, we research the inelastic
tunneling process only. By solving the time-dependent Schr{\"o}dinger
equation, we derive the inelastic tunneling current as follows:
\begin{align}
I_{s,t} & =\frac{2\pi e}{\hbar}\sum_{\nu,\nu^{\prime}}\int_{-\infty}^{\mu_{0}}dE_{n}\int_{\mu_{0}}^{0}d\xi_{k}\rho_{s}\left(E_{n}\right)\rho_{t}\left(\xi_{k}\right)\mathcal{N}_{s,t}^{2}\mid_{E_{n}\rightarrow\xi_{k}}\nonumber \\
 & \times\left\langle \nu\right|\nu^{\prime}\rangle^{2}\delta\left[E_{eg}+\left(\nu^{\prime}-\nu\right)\hbar\omega+\xi_{k}+eV_{b}-E_{n}\right],\label{eq:current_FC}
\end{align}
where $\mu_{0}$ is the Fermi level of electrodes, with the assumption
that the tip and the substrate are composed of the same metal. $\rho_{s}$
and $\rho_{t}$ denote the energy density of the substrate and the
tip, respectively. $E_{eg}$ is the energy gap of the molecule. $\mathcal{N}_{s,t}\mid_{E_{n}\rightarrow\xi_{k}}$
represents the inelastic tunneling matrix element describing the process
of the electrons tunneling from the substrate to the tip. It is written
as follows:
\begin{equation}
\mathcal{N}_{s,t}\mid_{E_{n}\rightarrow\xi_{k}}=-e\vec{\mu}\cdot\left\langle \phi_{k}\right|\frac{\vec{r}}{\left|\vec{r}\right|^{3}}\mid\varphi_{n}\rangle.
\end{equation}
Substituting the expression of the Franck-Condon factor into Eq. \ref{eq:current_FC},
we derive the expression of the inelastic current
\begin{align}
I_{s,t} & =\frac{2\pi e}{\hbar}\sum_{\nu^{\prime}}e^{-S}\frac{S^{\nu^{\prime}}}{\nu^{\prime}!}\int_{-\infty}^{\mu_{0}}dE_{n}\int_{\mu_{0}}^{0}d\xi_{k}\rho_{s}\left(E_{n}\right)\rho_{t}\left(\xi_{k}\right)\nonumber \\
 & \times\mathcal{N}_{s,t}^{2}\mid_{E_{n}\rightarrow\xi_{k}}\delta\left(E_{eg}+\nu^{\prime}\hbar\omega+\xi_{k}+eV_{b}-E_{n}\right).\label{eq:inelastic_current}
\end{align}
This expression can be simplified due to the $\delta$ function. For
convenience, we introduce the following definition: 
\begin{equation}
\Delta=E_{eg}+\nu^{\prime}\hbar\omega+eV_{b}.
\end{equation}
Thus, the range of the integral alters with the value of $\Delta$.
When $\mu_{0}<\Delta<0$, the current is written as 
\begin{align}
I_{s,t} & =\frac{2\pi e}{\hbar}\sum_{\nu^{\prime}}e^{-S}\frac{S^{\nu^{\prime}}}{\nu^{\prime}!}\int_{\mu_{0}}^{\mu_{0}-\Delta}d\xi_{k}\rho_{s}\left(\xi_{k}+\Delta\right)\rho_{t}\left(\xi_{k}\right)\mathcal{N}_{s,t}^{2}\mid_{\xi_{k}+\Delta\rightarrow\xi_{k}},\label{eq:current_delta_less_0}
\end{align}
When $\Delta<\mu_{0}$, the current becomes 
\begin{align}
I_{s,t} & =\frac{2\pi e}{\hbar}\sum_{\nu^{\prime}}e^{-S}\frac{S^{\nu^{\prime}}}{\nu^{\prime}!}\int_{\mu_{0}}^{0}d\xi_{k}\rho_{s}\left(\xi_{k}+\Delta\right)\rho_{t}\left(\xi_{k}\right)\mathcal{N}_{s,t}^{2}\mid_{\xi_{k}+\Delta\rightarrow\xi_{k}}.\label{eq:current_delta_less_mu0}
\end{align}

In experiments, both the elastic and the inelastic mechanism contribute
to the current spectrum. We can obtain the inelastic component via
the luminescence from the molecule. Once the molecule is excited to
its excited state, it then decays back through the spontaneous emission
process. We denote the photon-counting rate as $p_{e}\left(t\right)$.
The master equation of the molecule excited state reads \cite{Dong_2020}
\begin{equation}
\frac{dp_{e}\left(t\right)}{dt}=-\gamma p_{e}\left(t\right)+\frac{I_{s,t}}{e},
\end{equation}
in which $\gamma$ is the spontaneous decay rate.In the steady state,
the photon-counting rate becomes
\begin{equation}
\Gamma=\frac{I_{s,t}}{e}.
\end{equation}
Thus the inelastic tunneling current can be obtained from the photon-counting
spectrum.

\section{Results\label{sec:Results}}

\subsection{The inelastic tunneling current and the differential conductance}

For numerical calculations of the inelastic tunneling current, we
select the silver (Ag) as the material for both the tip and the substrate.
The Fermi level of silvers is $-4.64eV$. The characteristic frequency
of the molecular vibration is assumed to be $0.1eV$. The distance
between the tip and the substrate is set to $d=0.5$nm and the radius
of the apex atom of the tip is $0.5$nm. The tip is placed above the
molecule. Taking the symmetry of the system into account, we express
the position of the tip's apex as $\vec{a}=\left(0,0,d+R\right)$.
Furthermore, we choose different vibrational quantum numbers for the
electronic excited state, ranging from $0$ to $10$. We also consider
various cases with $S$ equal to $0.1$, $1$ and $10$, indicating
weak coupling, medium coupling and strong coupling between the vibration
and electrons, respectively.

The Fig. \ref{Fig.1.inelastic_current} illustrates the behavior of
the inelastic tunneling current (top figures) and the differential
conductance (bottom figures) varying with respect to the bias voltage.
The inelastic tunneling current increases as the bias voltage increases.
The onset bias voltage differs among Fig. 
\ref{Fig.1.inelastic_current} (a), (b) and (c), because the vibronic coupling alters the molecular
energy gap. Due to the Franck-Condon principle, a transition from
one vibrational energy level to another is more likely to occur if
the two vibrational wave functions overlap significantly. The effective
bias voltage is given by $eV_{eff}=-E_{eg}-\left[S\right]\hbar\omega$,
where $\left[S\right]$ is the floor function and expressed as $\left[S\right]=\max\left\{ n\in\mathbb{Z}\mid n\leq S\right\} $.The
curves of the differential conductance in Fig. \ref{Fig.1.inelastic_current}
(d), (e) and (f) exhibit steps structures. The spacing between these
steps corresponds to the characteristic frequency $\hbar\omega$ of
the molecular vibration.

{\footnotesize{}\vspace*{4mm}
}{\footnotesize\par}

{\footnotesize{}}%\centerline{\includegraphics{Fig.-1(file).eps}}}{\footnotesize\par}
\begin{center}
{\footnotesize{}}%
\parbox[c]{15.5cm}{%
\begin{center}
\includegraphics[width=15cm]{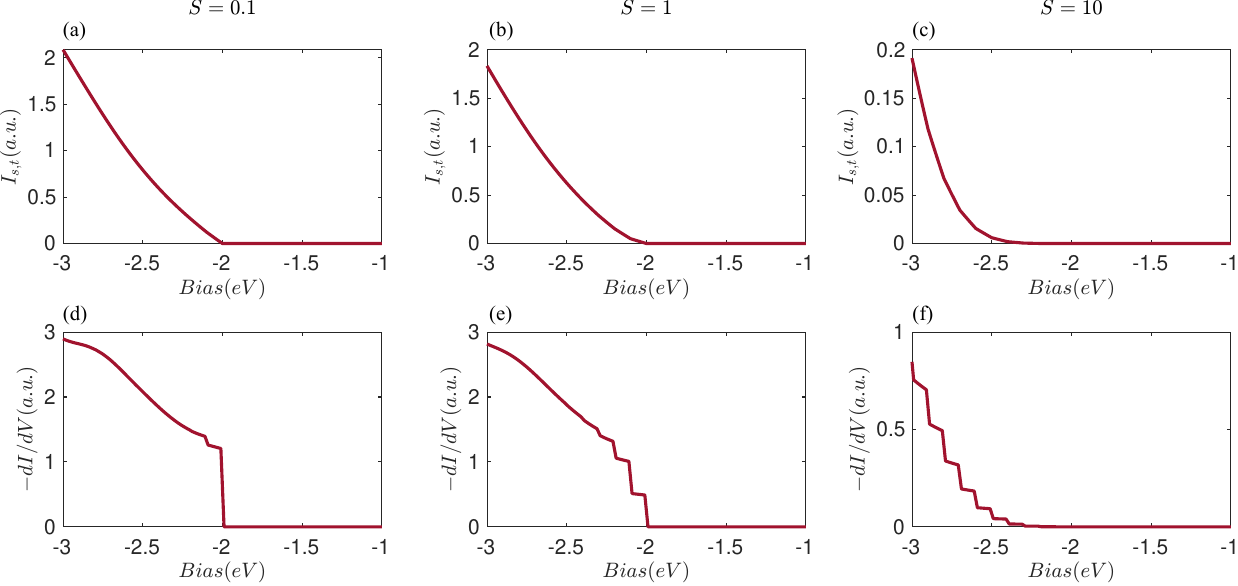}
\par\end{center}
\textbf{\small{}Fig.1.\label{Fig.1.inelastic_current}}{\small{} (a)-(c)
The inelastic tunneling current as a function of the bias voltage
with $S=0.1$, $S=1$ and $S=10$, respectively. (d)-(f) The differential
conductance as a function of the bias voltage with $S=0.1$,
$S=1$ and $S=10$, respectively. The characteristic frequency of
the molecular vibration is fixed at $0.1$eV.}%
}{\footnotesize\par}
\par\end{center}

{\footnotesize{}\vspace*{4mm}
}{\footnotesize\par}

\subsection{Determining the Huang-Rhys factor}

The red lines in Fig. \ref{Fig.2.second_derivative} illustrate the
normalized second derivative of the inelastic tunneling current with
respect to the bias voltage. Fig. \ref{Fig.2.second_derivative} (a),
(b) and (c) correspond to the parameter with $S=0.1$, $S=1$ and
$S=10$, respectively. These peaks in the graph represent the contribution
of the vibration to the inelastic tunneling current. The blue circles
represent the normalized values of the Franck-Condon factor for integer
values of $\nu$ varying from $0$ to $10$. The blue line connects
these circles in sequence. For small values of $\nu$, such as $\nu=0$
and $\nu=1$, the intensity of these peaks matches with blue circles
and the envelope of these peaks is consistent with the blue line very
well. Thus, for a given Huang-Rhys factor, the Franck-Condon factor
effectively presents the character of the intensity of the second
derivative of the inelastic tunneling current concerning the bias
voltage.

{\footnotesize{}\vspace*{4mm}
}{\footnotesize\par}

{\footnotesize{}}%\centerline{\epsfig{file=Fig-2.eps,width=7cm,height=8cm,clip=}}}{\footnotesize\par}
\begin{center}
{\footnotesize{}}%
\parbox[c]{15.5cm}{%
\begin{center}
\includegraphics[width=15cm]{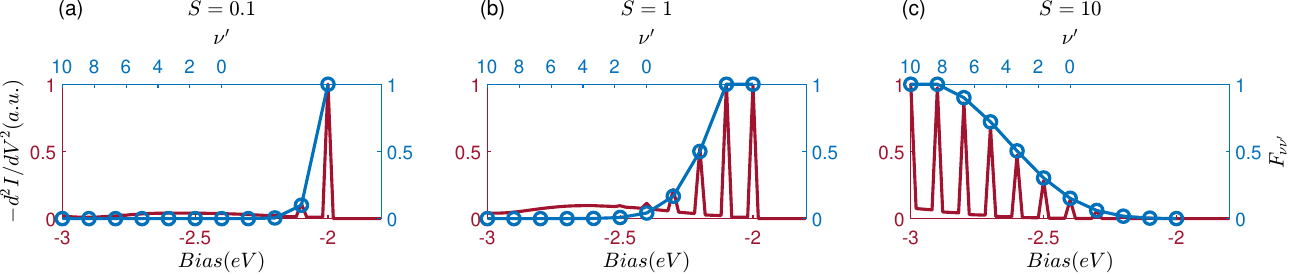}
\par\end{center}
\textbf{\small{}Fig.2.\label{Fig.2.second_derivative}}{\small{} The
second derivative of the inelastic tunneling current with respect
to the bias voltage (red) and the Frank-Condon factor (blue) as a
function of the bias voltage. (a)-(c) correspond to the case of $S=0.1$,
$S=1$ and $S=10$, respectively.}
}{\footnotesize\par}
\par\end{center}

{\footnotesize{}\vspace*{4mm}
}{\footnotesize\par}

From the expression of the inelastic tunneling current in Eq. (\ref{eq:inelastic_current}),
we can decompose $d^{2}I_{s,t}/dV^{2}$ with the vibronic quantum
number $\nu^{\prime}$, and define $\left(d^{2}I_{s,t}/dV^{2}\right)_{\nu^{\prime}}$
of the $\nu^{\prime}$-th order as 
\begin{align}
\left(d^{2}I_{s,t}/dV^{2}\right)_{\nu^{\prime}} & =e^{-S}\frac{S^{\nu^{\prime}}}{\nu^{\prime}!}D_{\nu^{\prime}}.\label{eq:d2I_nu}
\end{align}
Here, we have the expression of the factor $D_{\nu^{\prime}}$,
\begin{align}
D_{\nu^{\prime}} & =\frac{2\pi e}{\hbar}\frac{d^{2}}{dV^{2}}\int_{-\infty}^{\mu_{0}}dE_{n}\int_{\mu_{0}}^{0}d\xi_{k}\rho_{s}\left(E_{n}\right)\rho_{t}\left(\xi_{k}\right)\nonumber \\
 & \times\mathcal{N}_{s,t}^{2}\mid_{E_{n}\rightarrow\xi_{k}}\delta\left(E_{eg}+\nu^{\prime}\hbar\omega+\xi_{k}+eV_{b}-E_{n}\right).
\end{align}
 When the factor $D_{\nu^{\prime}}$ changes with $\nu^{\prime}$
as a slowly varying function, we get an approximate result about the
$\nu^{\prime}$-th order and the $\left(\nu^{\prime}-1\right)$-th
order of $d^{2}I_{s,t}/dV^{2}$ as
\begin{equation}
\frac{\left(d^{2}I_{s,t}/dV^{2}\right)_{\nu^{\prime}}}{\left(d^{2}I_{s,t}/dV^{2}\right)_{\nu^{\prime}-1}}\approx\frac{S}{\nu^{\prime}},
\end{equation}
especially\cite{Xu_2002,Xu2019} 
\begin{equation}
\frac{\left(d^{2}I_{s,t}/dV^{2}\right)_{1}}{\left(d^{2}I_{s,t}/dV^{2}\right)_{0}}\approx S,\label{eq:1_0_S}
\end{equation}
The ratio of the intensity of the first-order $d^{2}I_{s,t}/dV^{2}$
to the zero order $d^{2}I_{s,t}/dV^{2}$ is approximately equal to
the Huang-Rhys factor. This conclusion provides a theoretical foundation
for our experimental determination of the Huang-Rhys factor: Once
the inelastic tunneling current is obtained, we can analyze its second
derivative with respect to the bias voltage., from which the oscillation
quantum number $\nu^{\prime}$ can be obtained. Also, the interval
between peaks can give the value of the vibronic frequency $\omega$.
Through intensity of these peaks with the first order and the zero
order, we calculate the Huang-Rhys factor.

To verify the validity of the method for determining the Huang-Rhys
factor through the second derivative inelastic tunneling current, Fig.
\ref{Fig.3.estimated_true}illustrates the results estimated from
Eq.( \ref{eq:1_0_S}) as a function of the true value of $S$. The
red line is linear and represent the true value of $S$. The blue
circles are the estimated values from Eq. \ref{eq:1_0_S} with different
vibronic coupling strength , corresponding cases of $S=0.1$, $S=0.5$,
$S=1$, $S=5$ and $S=10$, respectively. There is a slight difference
between the estimated value and the true values in the case of weak
vibronic coupling. However, in the cases of the strong vibronic coupling,
the estimated $S$ matches the true $S$ very well. This figure shows
that $D_{\nu^{\prime}}$ is a slowly varying function, and Eq. \ref{eq:1_0_S}
can be used to determine the Huang-Rhys factor of the molecule.

{\footnotesize{}\vspace*{4mm}
}{\footnotesize\par}

{\footnotesize{}}%\centerline{\epsfig{file=Fig-2.eps,width=7cm,height=8cm,clip=}}}{\footnotesize\par}
\begin{center}
{\footnotesize{}}%
\parbox[c]{15.5cm}{%
\begin{center}
\includegraphics[width=8cm]{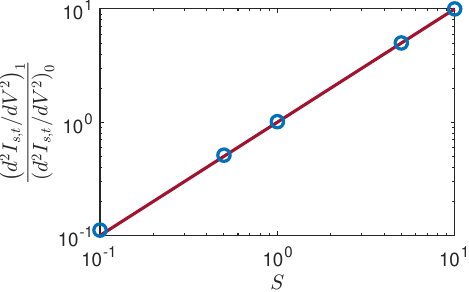}
\par\end{center}
\textbf{\small{}Fig.3.\label{Fig.3.estimated_true}}{\small{} The
comparison of the estimated value from Eq. \ref{eq:1_0_S} with the
true value. The red line is linear and represents the true value of
$S$. The blue circles are the estimated values from the left hand
of Eq. \ref{eq:1_0_S}, corresponding cases of $S=0.1$, $S=0.5$,
$S=1$, $S=5$ and $S=10$, respectively.}%
}{\footnotesize\par}
\par\end{center}

{\footnotesize{}\vspace*{4mm}
}{\footnotesize\par}

\section{Conclusion\label{sec:Conclusion}}

We have studied the STML system, where the molecule is modeled as
the two-level system with a molecular vibration degree of freedom. Based
on Bardeen's theory, the inelastic tunneling current is expressed
in terms of Huang-Rhys factor $S$. Information regarding vibronic
coupling is obtained from the differential conductance, which exhibits
steps as a function of the bias voltage. The Huang-Rhys factor is
determined from the second derivative of the inelastic tunneling current
with respect to the bias voltage. The method involving the analysis
of the inelastic tunneling current in STML will provide an alternative
approach for determining the Huang-Rhys factor.

\section*{Acknowledgments}

This work is supported by the National Natural Science Foundation
of China (NSFC) (Grant Nos. 11 875 049 and 12205211), the NSAF (Grant Nos. U1730449
and U1930403), and the National Basic Research Program of China (Grant
No. 2016YFA0301201).

{\small{}\baselineskip=10pt\itemsep-2pt}{\small\par}

{\small{}\bibliographystyle{unsrt}
\bibliography{HR}}

\begin{thebibliography}{10}

\bibitem{Qiu2003}
X.~H. Qiu, G.~V. Nazin, and W.~Ho.
\newblock Vibrationally resolved fluorescence excited with submolecular
  precision.
\newblock {\em Science}, 299(5606):542--546, jan 2003.

\bibitem{Schwarz_2015}
F.~Schwarz, Y.~F. Wang, W.~A. Hofer, R.~Berndt, E.~Runge, and J.~Kr{\"o}ger.
\newblock Electronic and vibrational states of single
  tin{\textendash}phthalocyanine molecules in double layers on {A}g(111).
\newblock {\em J. Phys. Chem. C}, 119(27):15716--15722, jun 2015.

\bibitem{Qiu_2004}
X.~H. Qiu, G.~V. Nazin, and W.~Ho.
\newblock Vibronic states in single molecule electron transport.
\newblock {\em Phys. Rev. Lett.}, 92(20):206102, may 2004.

\bibitem{Wu_2008}
S.~W. Wu, N.~Ogawa, G.~V. Nazin, and W.~Ho.
\newblock Conductance hysteresis and switching in a single-molecule junction.
\newblock {\em J. Phys. Chem. C}, 112(14):5241--5244, feb 2008.

\bibitem{Chen_2010}
Chi Chen, Ping Chu, C.~A. Bobisch, D.~L. Mills, and W.~Ho.
\newblock Viewing the interior of a single molecule: Vibronically resolved
  photon imaging at submolecular resolution.
\newblock {\em Phys. Rev. Lett.}, 105(21):217402, nov 2010.

\bibitem{Zhang_2016}
Yang Zhang, Yang Luo, Yao Zhang, Yun-Jie Yu, Yan-Min Kuang, Li~Zhang, Qiu-Shi
  Meng, Yi~Luo, Jin-Long Yang, Zhen-Chao Dong, and J.~G. Hou.
\newblock Visualizing coherent intermolecular dipole{\textendash}dipole
  coupling in real space.
\newblock {\em Nature}, 531(7596):623--627, mar 2016.

\bibitem{Komeda_2005}
Tadahiro Komeda.
\newblock Chemical identification and manipulation of molecules by vibrational
  excitation via inelastic tunneling process with scanning tunneling
  microscopy.
\newblock {\em Prog. Surf. Sci.}, 78(2):41--85, jan 2005.

\bibitem{Katano_2010}
Satoshi Katano, Sukekatsu Ushioda, and Yoichi Uehara.
\newblock Vibrational excitation of a single benzene molecule adsorbed on
  {C}u(110) studied by scanning tunneling microscope light emission
  spectroscopy.
\newblock {\em J. Phys. Chem. Lett.}, 1(19):2763--2768, sep 2010.

\bibitem{Jiang_2012}
N.~Jiang, E.~T. Foley, J.~M. Klingsporn, M.~D. Sonntag, N.~A. Valley, J.~A.
  Dieringer, T.~Seideman, G.~C. Schatz, M.~C. Hersam, and R.~P.~Van Duyne.
\newblock Observation of multiple vibrational modes in ultrahigh vacuum
  tip-enhanced raman spectroscopy combined with molecular-resolution scanning
  tunneling microscopy.
\newblock {\em Nano Lett.}, 12(10):5061--5067, jan 2012.

\bibitem{Zhang_2017}
Li~Zhang, Yun-Jie Yu, Liu-Guo Chen, Yang Luo, Ben Yang, Fan-Fang Kong, Gong
  Chen, Yang Zhang, Qiang Zhang, Yi~Luo, Jin-Long Yang, Zhen-Chao Dong, and
  J.~G. Hou.
\newblock Electrically driven single-photon emission from an isolated single
  molecule.
\newblock {\em Nat. Commun.}, 8(1), sep 2017.

\bibitem{Miwa_2019}
Kuniyuki Miwa, Hiroshi Imada, Miyabi Imai-Imada, Kensuke Kimura, Michael
  Galperin, and Yousoo Kim.
\newblock Many-body state description of single-molecule electroluminescence
  driven by a scanning tunneling microscope.
\newblock {\em Nano Lett.}, 19(5):2803--2811, jan 2019.

\bibitem{Kong_2021}
Fan-Fang Kong, Xiao-Jun Tian, Yang Zhang, Yun-Jie Yu, Shi-Hao Jing, Yao Zhang,
  Guang-Jun Tian, Yi~Luo, Jin-Long Yang, Zhen-Chao Dong, and J.~G. Hou.
\newblock Probing intramolecular vibronic coupling through vibronic-state
  imaging.
\newblock {\em Nat. Commun.}, 12(1), feb 2021.

\bibitem{Pe_a_Rom_n_2022}
Ricardo Javier~Pe{\~{n}}a Rom{\'{a}}n, Delphine Pommier, R{\'{e}}mi Bretel,
  Luis E.~Parra L{\'{o}}pez, Etienne Lorchat, Julien Chaste, Abdelkarim
  Ouerghi, S{\'{e}}verine~Le Moal, Elizabeth Boer-Duchemin, G{\'{e}}rald
  Dujardin, Andrey~G. Borisov, Luiz~F. Zagonel, Guillaume Schull,
  St{\'{e}}phane Berciaud, and Eric~Le Moal.
\newblock Electroluminescence of monolayer {WS} 2 in a scanning tunneling
  microscope: Effect of bias polarity on spectral and angular distribution of
  emitted light.
\newblock {\em Phys. Rev. B}, 106(8):085419, aug 2022.

\bibitem{Wen_2022}
Fei Wen, Guohui Dong, and Hui Dong.
\newblock Measuring fine molecular structures with luminescence signal from an
  alternating current scanning tunneling microscope.
\newblock {\em Commun. Theor. Phys.}, 74(12):125105, nov 2022.

\bibitem{Stipe1998}
B.~C. Stipe, M.~A. Rezaei, and W.~Ho.
\newblock Single-molecule vibrational spectroscopy and microscopy.
\newblock {\em Science}, 280(5370):1732--1735, jun 1998.

\bibitem{Stipe_1998}
B.~C. Stipe, M.~A. Rezaei, and W.~Ho.
\newblock Coupling of vibrational excitation to the rotational motion of a
  single adsorbed molecule.
\newblock {\em Phys. Rev. Lett.}, 81(6):1263--1266, aug 1998.

\bibitem{Lauhon_1999}
L.~J. Lauhon and W.~Ho.
\newblock Single-molecule vibrational spectroscopy and
  microscopy:{\hspace{0.6em}}{CO} on {C}u(001) and {C}u(110).
\newblock {\em Phys. Rev. B}, 60(12):R8525--R8528, sep 1999.

\bibitem{Lorente_2000}
N.~Lorente and M.~Persson.
\newblock Theory of single molecule vibrational spectroscopy and microscopy.
\newblock {\em Phys. Rev. Lett.}, 85(14):2997--3000, Oct 2000.

\bibitem{Pradhan_2005}
Nilay~A. Pradhan, Ning Liu, and Wilson Ho.
\newblock Vibronic spectroscopy of single {C}60 molecules and monolayers with
  the {STM}.
\newblock {\em J. Phys. Chem. B}, 109(17):8513--8518, feb 2005.

\bibitem{Nazin_2005}
G.~V. Nazin, S.~W. Wu, and W.~Ho.
\newblock Tunneling rates in electron transport through double-barrier
  molecular junctions in a scanning tunneling microscope.
\newblock {\em Proc. Natl. Acad. Sci.}, 102(25):8832--8837, jun 2005.

\bibitem{Huang1950}
Kun Huang, Avril Rhys, and Nevill~Francis Mott.
\newblock Theory of light absorption and non-radiative transitions in
  {F}-centres.
\newblock {\em Proc. R. Soc. Lond. A}, 204(1078):406--423, 1950.

\bibitem{Lemos_1965}
Anthony~M. Lemos and Jordan~J. Markham.
\newblock {C}alculation of the {H}uang-{R}hys factor for {F}-centers.
\newblock {\em J. Phys. Chem. Solids}, 26(12):1837--1851, dec 1965.

\bibitem{MULAZZI_1967}
E.~Mulazzi and Nice Terzi.
\newblock Evaluation of the {H}uang-{R}hys factor and the half-width of the
  {F}-band in {K}{C}l and {Na}{C}l crystals.
\newblock {\em J. Phys. Colloques}, 28(C4), aug 1967.

\bibitem{Moreno1992}
M~Moreno, M~T Barriuso, and J~A Aramburu.
\newblock The {H}uang-{R}hys factor {S}(a1g) for transition-metal impurities: a
  microscopic insight.
\newblock {\em J. Phys.: Condens. Matter}, 4(47):9481, nov 1992.

\bibitem{Schenk_1992}
A.~Schenk.
\newblock A model for the field and temperature dependence of
  {S}hockley-{R}ead-{H}all lifetimes in silicon.
\newblock {\em Solid-State Electron.}, 35(11):1585--1596, nov 1992.

\bibitem{Jong2015}
Mathijs de~Jong, Luis Seijo, Andries Meijerink, and Freddy~T. Rabouw.
\newblock Resolving the ambiguity in the relation between {S}tokes shift and
  {H}uang{\textendash}{R}hys parameter.
\newblock {\em Phys. Chem. Chem. Phys. PCCP}, 17(26):16959--16969, 2015.

\bibitem{Mingo_2000}
N.~Mingo and K.~Makoshi.
\newblock Calculation of the inelastic scanning tunneling image of acetylene on
  {C}u(100).
\newblock {\em Phys. Rev. Lett.}, 84(16):3694--3697, apr 2000.

\bibitem{Tikhodeev_2001}
S.~Tikhodeev, M.~Natario, K.~Makoshi, T.~Mii, and H.~Ueba.
\newblock Contribution to a theory of vibrational scanning tunneling
  spectroscopy of adsorbates.
\newblock {\em Surf. Sci.}, 493(1-3):63--70, nov 2001.

\bibitem{Mii_2002}
Takashi Mii, Sergei Tikhodeev, and Hiromu Ueba.
\newblock Theory of vibrational tunneling spectroscopy of adsorbates on metal
  surfaces.
\newblock {\em Surf. Sci.}, 502-503:26--33, apr 2002.

\bibitem{Hurley_2011}
Aaron Hurley, Nadjib Baadji, and Stefano Sanvito.
\newblock Spin-flip inelastic electron tunneling spectroscopy in atomic chains.
\newblock {\em Phys. Rev. B}, 84(3):035427, jul 2011.

\bibitem{Eickhoff_2020}
Fabian Eickhoff, Elena Kolodzeiski, Taner Esat, Norman Fournier, Christian
  Wagner, Thorsten Deilmann, Ruslan Temirov, Michael Rohlfing, F.~Stefan Tautz,
  and Frithjof~B. Anders.
\newblock Inelastic electron tunneling spectroscopy for probing strongly
  correlated many-body systems by scanning tunneling microscopy.
\newblock {\em Phys. Rev. B}, 101(12):125405, mar 2020.

\bibitem{Dong_2020}
Guohui Dong, Yining You, and Hui Dong.
\newblock Microscopic origin of molecule excitation via inelastic electron
  scattering in scanning tunneling microscope.
\newblock {\em New J. Phys.}, 22(11):113010, nov 2020.

\bibitem{Dong_2021}
Guohui Dong, Zhubin Hu, Xiang Sun, and Hui Dong.
\newblock Structural reconstruction of optically invisible state in a single
  molecule via scanning tunneling microscope.
\newblock {\em J. Phys. Chem. Lett.}, 12(41):10034--10039, oct 2021.

\bibitem{Drakova_1997}
D.~Drakova and G.~Doyen.
\newblock Local charge injection in {STM} as a mechanism for imaging with
  anomalously high corrugation.
\newblock {\em Phys. Rev. B}, 56(24):R15577--R15580, dec 1997.

\bibitem{Galperin_2005}
Michael Galperin and Abraham Nitzan.
\newblock Current-induced light emission and light-induced current in
  molecular-tunneling junctions.
\newblock {\em Phys. Rev. Lett.}, 95(20):206802, nov 2005.

\bibitem{Harbola_2006}
Upendra Harbola, Jeremy~B. Maddox, and Shaul Mukamel.
\newblock Many-body theory of current-induced fluorescence in molecular
  junctions.
\newblock {\em Phys. Rev. B}, 73(7):075211, feb 2006.

\bibitem{Jiang_2023}
Song Jiang, Tom{\'{a}}{\v{s}} Neuman, R{\'{e}}mi Bretel, Alex Boeglin, Fabrice
  Scheurer, Eric~Le Moal, and Guillaume Schull.
\newblock Many-body description of {STM}-induced fluorescence of charged
  molecules.
\newblock {\em Phys. Rev. Lett.}, 130(12):126202, mar 2023.

\bibitem{Doppagne_2017}
Benjamin Doppagne, Michael~C. Chong, Etienne Lorchat, St{\'{e}}phane Berciaud,
  Michelangelo Romeo, Herv{\'{e}} Bulou, Alex Boeglin, Fabrice Scheurer, and
  Guillaume Schull.
\newblock Vibronic spectroscopy with submolecular resolution from {STM}-induced
  electroluminescence.
\newblock {\em Phys. Rev. Lett.}, 118(12):127401, mar 2017.

\bibitem{Mart_n_Jim_nez_2020}
Alberto Mart{\'{\i}}n-Jim{\'{e}}nez, Antonio~I.
  Fern{\'{a}}ndez-Dom{\'{\i}}nguez, Koen Lauwaet, Daniel Granados, Rodolfo
  Miranda, Francisco~J. Garc{\'{\i}}a-Vidal, and Roberto Otero.
\newblock Unveiling the radiative local density of optical states of a
  plasmonic nanocavity by {STM}.
\newblock {\em Nat. Commun.}, 11(1), feb 2020.

\bibitem{Zhu_2021}
Jia-Zhe Zhu, Gong Chen, Talha Ijaz, Xiao-Guang Li, and Zhen-Chao Dong.
\newblock Influence of an atomistic protrusion at the tip apex on enhancing
  molecular emission in tunnel junctions: A theoretical study.
\newblock {\em J. Chem. Phys.}, 154(21), jun 2021.

\bibitem{Miwa_2023}
Kuniyuki Miwa, Souichi Sakamoto, and Akihito Ishizaki.
\newblock Control and enhancement of single-molecule electroluminescence
  through strong light{\textendash}matter coupling.
\newblock {\em Nano Lett.}, 23(8):3231--3238, apr 2023.

\bibitem{Bardeen_1961}
J.~Bardeen.
\newblock {T}unnelling from a {M}any-{P}article {P}oint of {V}iew.
\newblock {\em Phys. Rev. Lett.}, 6(2):57--59, jan 1961.

\bibitem{Franck_1926}
J.~Franck and E.~G. Dymond.
\newblock Elementary processes of photochemical reactions.
\newblock {\em Trans. Faraday Soc.}, 21(February):536, 1926.

\bibitem{Condon_1926}
Edward Condon.
\newblock {A} {T}heory of {I}ntensity {D}istribution in {B}and {S}ystems.
\newblock {\em Phys. Rev.}, 28(6):1182--1201, dec 1926.

\bibitem{Xu_2002}
S.~J. Xu, W.~Liu, and M.~F. Li.
\newblock Direct determination of free exciton binding energy from
  phonon-assisted luminescence spectra in {GaN} epilayers.
\newblock {\em Appl. Phys. Lett.}, 81(16):2959--2961, oct 2002.

\bibitem{Xu2019}
Shi-Jie Xu.
\newblock {H}uang-{R}hys factor and its key role in the interpretation of some
  optical properties of solids.
\newblock {\em Acta Phys. Sin-ch. Ed.}, 68(16):166301, 2019.

\end{thebibliography}
{\small\par}

%\end{CJK*}
\end{document}